%% file: paper-draft-epjc-v1.0.tex
\RequirePackage{fix-cm}

\RequirePackage{lineno} 

\documentclass[twocolumn,epjc3]{svjour3}  
\smartqed

\RequirePackage{graphicx}

\usepackage{preprintcover}
\PreprintCoverPaperTitle{Search for resonant diboson production in the $\ellell\qqbar$ final state in $pp$ collisions at $\sqrt{s} = 8$~TeV with the ATLAS detector}
\PreprintIdNumber{CERN-PH-EP-2014-211}  
\PreprintCoverAbstract{This paper reports on a search for narrow resonances in diboson production in the $\ellell\qqbar$ final state 
using $pp$ collision data corresponding to an integrated luminosity of $20$~fb$^{-1}$ 
collected at $\rts=8\TeV$ with the
ATLAS detector at the Large Hadron Collider. No significant excess of data events over the Standard Model expectation is
observed. Upper limits at the 95\% confidence level are set on the production cross section 
times branching ratio for Kaluza--Klein gravitons predicted by the Randall--Sundrum model
and for Extended Gauge Model $W'$ bosons. These results lead to the exclusion of mass values
below 740~GeV and 1590~GeV for the graviton and $W'$ boson respectively.}
\PreprintJournalName{EPJC}

\usepackage{epstopdf}
\usepackage{subfigure}
\usepackage{amsmath}
\usepackage{amssymb}
\usepackage{mathrsfs}
\usepackage{collref}
\usepackage{stfloats} 
\usepackage{float} 
\usepackage{url}
\usepackage{cite}
\usepackage{rotating}
\usepackage{multirow}

\input{definitions}

\journalname{Eur. Phys. J. C}

\begin{document}

\title{Search for resonant diboson production in the $\ellell\qqbar$ final state in $pp$ collisions at $\sqrt{s} = 8$~TeV with the ATLAS detector}

\author{ATLAS Collaboration{}}
\institute{\label{}}

\date{Received: date / Accepted: date}

\maketitle

\begin{abstract}
This paper reports on a search for narrow resonances in diboson production in the $\ellell\qqbar$ final state 
using $pp$ collision data corresponding to an integrated luminosity of $\lumiNoDecimal$~fb$^{-1}$ 
collected at $\rts=8\TeV$ with the
ATLAS detector at the Large Hadron Collider. No significant excess of data events over the Standard Model expectation is
observed. Upper limits at the 95\% confidence level are set on the production cross section 
times branching ratio for Kaluza--Klein gravitons predicted by the Randall--Sundrum model
and for Extended Gauge Model $\Wprime$ bosons. These results lead to the exclusion of mass values
below \GravitonMassLimitObs{}~GeV and \WprimeMassLimitObs~GeV for the graviton and $\Wprime$ boson respectively.

\end{abstract}

\section{Introduction}
\label{intro}
This paper presents a search for narrow diboson resonances in the
semileptonic decay channel $\Z\W$ or $\Z\Z\rightarrow \ellell\qqbar$ (where $\ell$ stands for electron or muon) in
$pp$ collision data  corresponding to an integrated luminosity of \lumiNoDecimal\;\ifb\;
recorded with the ATLAS
detector at a centre-of-mass energy $\rts=8\TeV$ at the Large Hadron Collider (LHC). 
This type of resonances appear in models 
such as Technicolor~\cite{tc1,tc2,tc3}, Warped Extra
Dimensions~\cite{rs1,rs2,rs3}, and Grand Unified Theories~\cite{GUT1,GUT2,GUT3,GUT4}. 
The semileptonic decay channel has  
a relatively large branching ratio compared to the fully leptonic mode, while
the requirement of the presence of two decay leptons can 
suppress the multijet background present in the fully hadronic mode. Additionally, the 
absence of neutrinos in the final state allows to reconstruct the invariant mass of the
diboson system. 

This analysis is optimized using two models with narrow resonances as benchmarks:
spin-2 Kaluza--Klein (KK) gravitons ($\Graviton \rightarrow \Z\Z$), and 
spin-1 \Wprime{} gauge bosons ($\Wprime \rightarrow \Z\W$) of the
Sequential Standard Model (SSM) with modified coupling to $\Z\W$, 
also referred to as the Extended Gauge Model (EGM)~\cite{EGM}.
For both models, the $\W$ and $\Z$ bosons from resonance decays are longitudinally 
polarized over a pole mass range relevant to this analysis.

The KK graviton interpretation is based on an extended Randall--Sundrum (RS) model  
with a warped extra dimension in which the Standard Model (SM) fields can propagate~\cite{bulkrs}.
This extended ``bulk'' RS model avoids
constraints on the original RS model~\cite{rs1}, referred to as RS1 hereafter, 
from limits on flavour-changing neutral currents and from electroweak precision tests.
The bulk RS model is characterized by a dimensionless coupling constant $k/\bar{M}_{\rm Pl} \sim 1$
where $k$ is the curvature of the warped extra dimension and 
$\bar{M}_{\rm Pl}=M_{\rm Pl}/\sqrt{8\pi}$ is the reduced Planck mass.
The width relative to the mass of the bulk RS graviton with $k/\bar{M}_{\rm Pl}=1$ 
varies between 3\% and 6\% within the pole mass range of 300--2000~GeV.

The EGM introduces $W'$ and $Z'$ bosons with SM couplings to 
fermions and with the coupling strength of the heavy $W'$ to $WZ$ modified
by a mixing factor $\xi = c \times (\mW/\mWprime)^2$ relative to the SM couplings,	 
where $\mW$ and $\mWprime$ are the pole masses of the \W{} and \Wprime{} bosons respectively, and 
$c$ is a coupling scaling factor. In this scenario the partial width of the \Wprime{} boson to $WZ$ 
scales linearly with \mWprime{}, leading to a narrow resonance over 
the accessible mass range, in contrast to the SSM where the width
grows rapidly as $\mWprime^5$. For the simulated EGM $W'$ samples used in the analysis, 
the natural \Wprime{} width is about 3\% at a pole mass of 300~GeV and increases slightly to 4\%  at a pole mass of 2000~GeV.

Previous searches for diboson resonances have been carried out using $\ppbar$ collision data at $\rts=1.96\TeV$ 
at the Tevatron and $pp$ collision data at $\rts =$~7--8~TeV at the LHC.
The \dzero{} Collaboration
searched for resonances in \WW{} and \Z\W{} production~\cite{D01,D02} and 
excluded \Wprime{} bosons in the mass range of 180--690~GeV 
and RS1 gravitons in the mass interval 300--754~GeV at the 95\% confidence level (CL).  The CDF experiment 
searched for resonances in the \Z\Z{} decay channel
and set limits on the production cross section of RS1 gravitons in the mass range 300--1000~GeV 
at the 95\% CL~\cite{CDF3}.
The ATLAS Collaboration reported searches for resonant \ZZ{} ($\to \ell\ell\ell'\ell'$, $\ell\ell\qqbar$){}~\cite{ATLAS_ZZ4l_2l2j}, 
\WW{} ($\to \ell\nu\qqbar$)~\cite{ATLAS_WWlvjj} and \WZ{} ($\to \ell\nu\qqbar$, $\ell\nu\ell'\ell'$)~\cite{ATLAS_WWlvjj,ATLAS_WZlvll} 
production, and searches for
new phenomena in high-mass \WW{} ($\to \ell\nu\ell'\nu'$) processes~\cite{ATLAS_WWlvlv}
using $pp$ collision data recorded at $\sqrt{s}=7$~TeV, except for the $\WZ \to \ell\nu\ell'\ell'$ search 
in Ref.~\cite{ATLAS_WZlvll}, which used data recorded at $\sqrt{s}=8$~TeV. Here $\ell'$ stands for an electron or muon.
These studies excluded EGM \Wprime{} bosons with masses up to 1.52~TeV for $\WZ$ final states, 
RS1 gravitons with masses up to 845~GeV for \ZZ{} final states and up to 1.23~TeV for $\WW$ final states.
The CMS Collaboration searched for \Z\Z{} and \W\W{} resonances in the semileptonic decay 
channel, setting exclusion limits on the production cross section of bulk RS gravitons~\cite{CMS-PAS-EXO-12-022}.  
In the fully hadronic channel, the CMS Collaboration excluded RS1 gravitons with $k/\bar{M}_{\rm Pl}$=0.1 for masses up to 1.2~TeV, 
and $W'$ bosons for masses up to 1.7~TeV~\cite{CMS-PAS-EXO-12-024}. Both of these searches implement jet substructure
techniques to identify the event topology where the hadronic system from the decay of one or two gauge bosons 
is produced at high transverse momentum $\pT$, resulting in a single reconstructed jet. 
In the \mbox{analysis} presented here, a similar technique has been used to identify hadronically decaying $W$ or $Z$ boson 
produced at high $\pt$. This technique uses the characteristics of two cores (``subjets'') 
inside a single reconstructed jet and allows for a significant improvement in acceptance and selection efficiency  
for high mass states with boosted $W$ and $Z$ bosons over the previous analysis~\cite{ATLAS_ZZ4l_2l2j}.

\section{Analysis}
\label{sec:selection}
In this study, three optimized sets of selection criteria classify $\Z\W/\Z\Z \rightarrow \ellell\qqbar$ events 
into distinct kinematic regions, namely the ``low-\pT{} resolved region'' (LR), 
``high-\pT{} resolved region'' (HR) and ``merged region'' (MR), based on the $\pT$ of the dilepton and the hadronic system.  
In the LR and HR the hadronic boson decay is reconstructed as two distinct jets, whereas in the 
MR it is reconstructed as a single jet. 
In all three cases, the dilepton (hadronic system) mass is required to be consistent
with the mass of the $\Z$ boson ($\W$ or $\Z$ boson).
In the MR, additional jet substructure information, optimized for the identification of the hadronic decay of a longitudinally 
polarized high-$\pT$ boson, is used to improve the sensitivity.
Finally, the $\ellell\qqbar$ mass spectrum, reconstructed as the mass of the dilepton and the two-jet system in the LR and HR ($m_{\ell\ell jj}$) or the dilepton and the single-jet system in the MR ($m_{\ell\ell J}$), is examined for excesses with respect 
to the expectation from SM processes (background).

\subsection{Detector and data sample}
\label{sec:ATLAS}
The ATLAS detector~\cite{atlas} consists of an inner detector (ID) providing charged particle tracking for the 
pseudorapidity\footnote{ATLAS uses a right-handed coordinate system with the $z$-axis along the beam pipe. The $x$-axis points to the centre of the LHC ring, and the $y$-axis points upward. Cylindrical coordinates ($R$,$\phi$) are used in the transverse plane, $\phi$ being the azimuthal angle around the beam pipe. The pseudorapidity is defined in terms of the polar angle $\theta$ as $\eta=-\ln(\tan(\theta/2))$.}
range $|\eta|<2.5$, surrounded by a superconducting solenoid, 
electromagnetic and hadronic calorimeters with a coverage of $|\eta| < 4.9$, and a muon spectrometer 
(MS) with toroidal magnets that provides muon identification in the range $|\eta|<2.7$. 

This study uses an integrated luminosity of $\lumi~\ifb$ of $pp$ collision data collected in
2012. The luminosity is derived from beam-separation scans~\cite{atlaslumi2011} 
and has an uncertainty of \lumiRelErr\%.
Events are selected with lepton triggers that require the presence of at least one 
lepton (electron or muon) with \pT{} above 24~GeV.
The trigger efficiency for signal events that pass the 
selection criteria described in Sect.~\ref{sec:object} 
is approximately 92\% for the muon channel and greater than 99\% for the electron channel. 

\subsection{Simulated event samples}
\label{sec:MC}
To model the acceptance and the reconstructed mass spectra for narrow resonances, 
benchmark signal samples are generated with pole masses between 300 and 2000~GeV, in 100~GeV steps.
Additional samples are generated between 350 and 950~GeV for the bulk RS $\Graviton$ signal so that 
the mass gap is reduced to 50~GeV, which is comparable to the detector resolution of the reconstructed $\ellell\qqbar$
mass in this mass interval.
The bulk RS $\Graviton$ signal events are generated by \calchep~\cite{b-calchep}
with $k/\bar{M}_{\rm Pl}=1.0$, and the \Wprime{} signal sample is generated with
\pythia8~\cite{b-pythia}, setting the coupling scale factor $c=1$. The factorization and renormalization scales 
are set to the resonance mass. 
The hadronisation and fragmentation are modelled with \pythia8 in both 
cases, and the CTEQ6L1~\cite{b-cteq61l} (MSTW2008LO~\cite{b-mstw2008lo}) 
parton distribution functions (PDFs) are used for the $\Graviton$ ($\Wprime$) signal.
The $\Wprime$ production cross section is scaled to a next-to-next-to-leading-order (NNLO) calculation
in $\alpha_s$ by \ZWPROD\cite{b-zwprod}.
Calculated production cross section times branching ratio values for different pole masses are given in 
Table~\ref{tab:xsecsTheo}.
\begin{table}
\begin{center}
  \caption{Theoretical production cross section times branching ratio
  $\sigma_{\Graviton} (pp \to \Graviton)\cdot{\rm BR}(\Graviton \to \Z\Z)$ with $k/\bar{M}_{\rm Pl}=1.0$ and 
  $\sigma_{\Wprime} (pp \to \Wprime)\cdot{\rm BR}(\Wprime \to \Z\W)$ with $c=1$ for different 
  pole masses of the resonances.
    \label{tab:xsecsTheo}}
 \begin{tabular}{c|c|c}
\hline\hline
Pole mass [GeV] & 
$\sigma_{\Graviton}\cdot{\rm BR}$~[fb] &
$\sigma_{\Wprime}\cdot{\rm BR}$~[fb]
\\ \hline
500  & 540   & 4100\\
800  & 23    &  540\\
1400 & 0.44  &   33\\
\hline
  \end{tabular}
\end{center}
\end{table}

The main background sources are \Z{} bosons produced in association with jets ($\Zjet$), followed
by top-quark pair and non-resonant vector-boson pair production.
The contribution from multijet events is negligible after the selection cuts described in Sect.~\ref{sec:object}.
All background estimates are based on simulation. Additionally, the main background source, 
$\Zjet$, is estimated using constraints from data as described in Sect.~\ref{sec:background}. 
The $\Zjet$ background is modelled by the
\sherpa generator~\cite{b-sherpa} with CT10 PDFs~\cite{b-ct10}. 
The top pair, $s$-channel single-top and $Wt$ processes are modelled by the \mcatnlo\cite{b-mcnlo}
generator with CT10 PDFs, interfaced to \HERWIG\cite{b-herwig} for hadronisation and \JIMMY\cite{b-jimmy} for 
modelling of the underlying event.
The top pair production cross section is calculated at NNLO in QCD including resummation 
of next-to-next-to-leading logarithmic soft gluon terms with Top++2.0~\cite{b-top1,b-top2,b-top3,b-top4,b-top5,b-top6}.
The $t$-channel single-top events are generated by AcerMC~\cite{b-acer} with CTEQ6L1 PDFs and \PYTHIA6~\cite{b-pythia6} for hadronisation.
The diboson events are produced with the \HERWIG generator and CTEQ6L1 PDFs. 
The diboson production cross sections are normalized to predictions at next-to-leading-order accuracy as calculated with \	\cite{b-mcfm}.
Generated events are processed with the ATLAS detector simulation program~\cite{b-atlassim} based on 
the GEANT4 package~\cite{b-geant4}. Effects from additional inelastic $pp$ interactions (pile-up) 
occurring in the same bunch crossing are taken into account by overlaying minimum-bias events 
simulated by \PYTHIA8.

\subsection{Object and event selection}
\label{sec:object}
Electron candidates are selected from energy clusters in the electromagnetic calorimeter according to the medium criteria 
of Ref.~\cite{b-electron-isem}, which impose requirements on the shower profile and demand 
an associated ID track.
{Offline} reconstructed electrons are required to have $\pT > 25$~GeV and $|\eta|$ $<$ 2.47. The transition region 
 between the barrel and endcaps ($1.37 < |\eta| < 1.52$) exhibits degraded energy resolution and is therefore excluded.

Muon candidates are reconstructed by combining ID and MS tracks which 
have consistent position, charge and momentum measurements~\cite{b-muon-reco}.  
The muon candidates are required to have $\pt>25$~GeV and $|\eta|<2.4$. 

A primary vertex reconstructed from at least three well-reconstructed charged
particle tracks, each with $\pT > 400\MeV$, is required in order to remove 
non-collision background.
If an event contains more than one primary vertex candidate, the vertex with the highest
$\Sigma\pT^2$ of the associated tracks is selected.
To ensure that both electrons or muons originate from the primary vertex,
it is required that the product of the longitudinal impact parameter ($z_0$) and 
the sine of the polar angle of the candidate ($\theta$) satisfies $|z_0 \sin(\theta)|<0.5$~mm, 
and that the ratio of the transverse impact parameter ($d_0$) to 
its uncertainty ($\sigma_{d_0}$) for electrons (muons) 
fulfils $|d_0|/\sigma_{d_0}<6$ (3.5).
In addition, the lepton candidates are required to be isolated from other tracks
and calorimetric activity. The scalar sum of the transverse momenta of tracks 
within a cone of size $\Delta R\equiv\sqrt{(\Delta\eta)^2+(\Delta\phi)^2}=0.2$ around the lepton track
is required to be less than 15\% of the candidate $\pT$. Similarly, the sum of transverse energy 
deposits in the calorimeter within a cone of size $\Delta R=0.2$, 
excluding the transverse energy due to the lepton, and corrected for the expected pile-up contribution,
is required to be less than 30\% of the candidate $\pT$ (calorimetric isolation). 

To improve the acceptance for events with boosted \Z{} bosons, with $\pT > 800$~GeV, 
the isolation method is modified for dilepton objects: 
a dilepton track isolation variable is calculated for each lepton of a 
like-flavour pair by subtracting the $\pT$ of the paired lepton from the $\pT$ sum described above if it falls inside the isolation cone of the lepton under consideration.
The modified scalar sum $\pT$ variable for the dilepton isolation is
required to be less than 15\% of the lepton $\pT$, as in the standard track isolation.
The calorimetric isolation requirements are dropped if $\Delta R (\ellell) < 0.25$. 

Jets are reconstructed from clusters of calorimeter cells using the anti-$k_t$ 
algorithm~\cite{b-antikt} with a distance parameter $R=0.4$.
Jets are required to be in the range $|\eta|<2.1$ and to have $\pT>30$~GeV after correcting for energy losses in passive material, the non-compensating response 
of the calorimeter and extra energy due to event pile-up~\cite{b-jes}.
Furthermore, for jets with $\pT<50$~GeV, the scalar sum $\pT$ of associated tracks 
from the primary vertex is required to be at least 
50\% of the scalar sum $\pT$ of all associated tracks to suppress jets from pile-up interactions.
The selected anti-$k_t$ jets are referred to as small-$R$ jets and denoted by ``$j$'' hereafter.

For resonances with a mass above about 900~GeV,
the $q\bar{q}$ pair is often merged into a single jet 
and the fraction of merged $q\bar{q}$ pairs increases with the resonance mass. 
Such jets are reconstructed with
the Cambridge--Aachen jet clustering algorithm~\cite{b-cambridge-aachen} with a distance parameter $R=1.2$.  
To exploit the characteristics of the decay of the massive boson into a $q\bar{q}$ pair, 
these jets are further required to pass a splitting and filtering algorithm similar to the
algorithm described in Ref.~\cite{b-butterworth} but optimized for the identification of very high-$\pT$ 
boson decays~\cite{b-boson-tagging-note}.  These jets are required to be within 
$|\eta|<1.2$ and to have $\pT>100$~GeV, and are referred to as large-$R$ jets or ``$J$'' hereafter.

Events which contain exactly two electrons or muons satisfying the above criteria are selected if at least one is associated with a lepton trigger candidate.
To select lepton pairs originating from a $Z$ boson decay,
the dilepton invariant mass ($m_{\ell\ell}$) is required to be 
in the range $66\GeV<m_{\ell\ell}<116$~GeV. 
The $m_{\ell\ell}$ cut range is chosen to be wide to enhance the signal sensitivity, given that 
the dominant background is from the $\Zjet$ processes and a narrower cut would not provide
additional discrimination power.
Muon-pair events are further required to have muons of opposite charge.
The opposite charge requirement is not required for electron-pair events because of a higher charge 
misidentification rate for high-$\pT$ electrons. 

The three selection regions are differentiated by the $\pT$ ranges for the leptonic 
$\Z$ decay candidate ($\pT^{\ell\ell}$)
and hadronic jet system, namely 
$\pT^{\ell\ell}>400$~GeV and $\pT^J>400$~GeV for the large-$R$ jet in the MR, and $\pT^{\ell\ell}>100$ (250)~GeV and 
$\pT^{jj}>100$ (250)~GeV for the two small-$R$ jets at low (high) $\pT$ in the LR (HR).  
The mass of the hadronic jet system is required to be in the range $70$~GeV $< m_{jj/J} < 110$~GeV for both the hadronic $\W$ and $\Z$ decay candidates in all three regions. 
In the MR, the large-$R$ jet is split into subjets using an 
algorithm described in Ref.~\cite{b-butterworth}. However, in contrast to the configuration used in Ref.~\cite{b-butterworth},
the mass relation between the large-$R$ jet and subjets, the mass drop, is not imposed.
A subjet momentum balance variable is defined as
 $\sqrt{y_{\mathrm f}} = \min(\pT^{j1}, \pT^{j2})\Delta R_{12}/m_{12}$, 
 where $\pT^{j1}$ and $\pT^{j2}$ are the transverse momenta of the two leading 
 subjets, $\Delta R_{12}$ is their separation and $m_{12}$ is their mass. 
 To suppress jets from gluon radiation and splitting, the subjet momentum balance is required to be $\sqrt{y_{\mathrm f}}>0.45$.
 Events are classified by sequentially applying the 
criteria for the MR, HR, and LR, thus 
assigning each event exclusively to one region.
Overall, the signal acceptance times efficiency after all selection requirements increases 
from 5--10\% at $m_{G^*}=$~300~GeV to a plateau of 30--35\% above $m_{G^*}=$~500~GeV
for a signal sample of $\Graviton \rightarrow \Z\Z\rightarrow \ellell\qqbar$.
The improvement in acceptance compared to the previous analysis \cite{ATLAS_ZZ4l_2l2j} ranges up 
to a factor of five for masses above 1.5 TeV.

\subsection{Background and event yield}
\label{sec:background}
The simulation of the main background source ($\Zjet$) is corrected using data.
The normalization and $m_{\ell\ell jj/J}$ shape corrections of the simulated $\Zjet$ background sample  
is determined from data in a control region defined by all selection cuts but with an inverted 
cut on $m_{jj/J}$, namely $m_{jj/J}<70$~GeV or $m_{jj/J}>110$~GeV, in the resolved and merged regions, respectively. 
The normalization corrections, obtained as the ratio of event yields 
in the data and $\Zjet$ simulated samples for the electron and muon channel, 
after removing contributions from subdominant backgrounds from the data spectrum,
range between 2\% and 10\%. 
The $m_{\ell\ell jj/J}$ shape correction is well reproduced with 
a linear fit to the ratio of data to $\Zjet$ background, derived for each signal region after 
combining the electron and muon channels. This results in bin-by-bin corrections of up to 7\% in the LR, 
3\% in the HR, and 22\% in the MR. 
The other backgrounds, from diboson and top production, are taken from simulation without
applying corrections from data control regions. 

The event yield in the three signal regions is summarized in Table~\ref{tab:yield}.  
The total event yield with combined statistical and systematic uncertainties is given before and after the 
simultaneous fit to the three signal regions (cf. Sect.~\ref{sec:result}).  

\begin{table*}
\begin{center}
  \hfill{}
  \caption{Event yields in signal regions for data, expected backgrounds, and $\Graviton$ and $\Wprime$ signals.
 The statistical and systematic uncertainties are given separately (in this order), except for the total background 
where the combined statistical and systematic uncertainty before (unconstrained) and after (constrained) the fit to 
the data in the signal regions is shown.
The signal mass points correspond to 500 (LR), 800 (HR), 1400 (MR) GeV.
\label{tab:yield}}
\begin{tabular*}{\textwidth}{c @{\extracolsep{\fill}} ccc} 
\hline\hline
  Sample                  &   LR                           &   HR                           &   MR                              \\ \hline
  $\Zjet$                 & 9460   $\pm$ 40   $\pm$  660   &  591    $\pm$ 4   $\pm$  15    &  20.9   $\pm$ 0.3    $\pm$ 2.3    \\
  $WW/WZ/ZZ$              &  234   $\pm$  4   $\pm$   22   &   20.6  $\pm$ 0.3 $\pm$  1.4   &   1.38  $\pm$ 0.02   $\pm$ 0.13  \\ 
  $t\bar{t}$ + Single $t$ &  175.3 $\pm$  9.2 $\pm$    9.9 &    -                           &   -                               \\ \hline
  Total (unconstrained)   & 9870   $\pm$ 690               &  612    $\pm$ 17               &  22.3   $\pm$ 2.5                \\
  Total (constrained)     & 9730   $\pm$  98               &  608.8  $\pm$  3.8             &  21.80  $\pm$ 0.46                \\ \hline
  Data                    & 9728                           &  619                           &  25                               \\ \hline
  $\Graviton$ Signal      & 1097   $\pm$ 17   $\pm$  63    &   14.27 $\pm$ 0.19 $\pm$ 0.76  &  0.0995 $\pm$ 0.0013 $\pm$ 0.0059 \\
  $W'$ Signal             & 1950   $\pm$ 40   $\pm$ 140    &  145.0  $\pm$ 2.3  $\pm$ 8.1   &  3.64   $\pm$ 0.06   $\pm$ 0.31   \\ 
  \hline\hline
\end{tabular*}
\end{center}
\end{table*}

\begin{figure}\centering
\includegraphics[width=0.8\columnwidth]{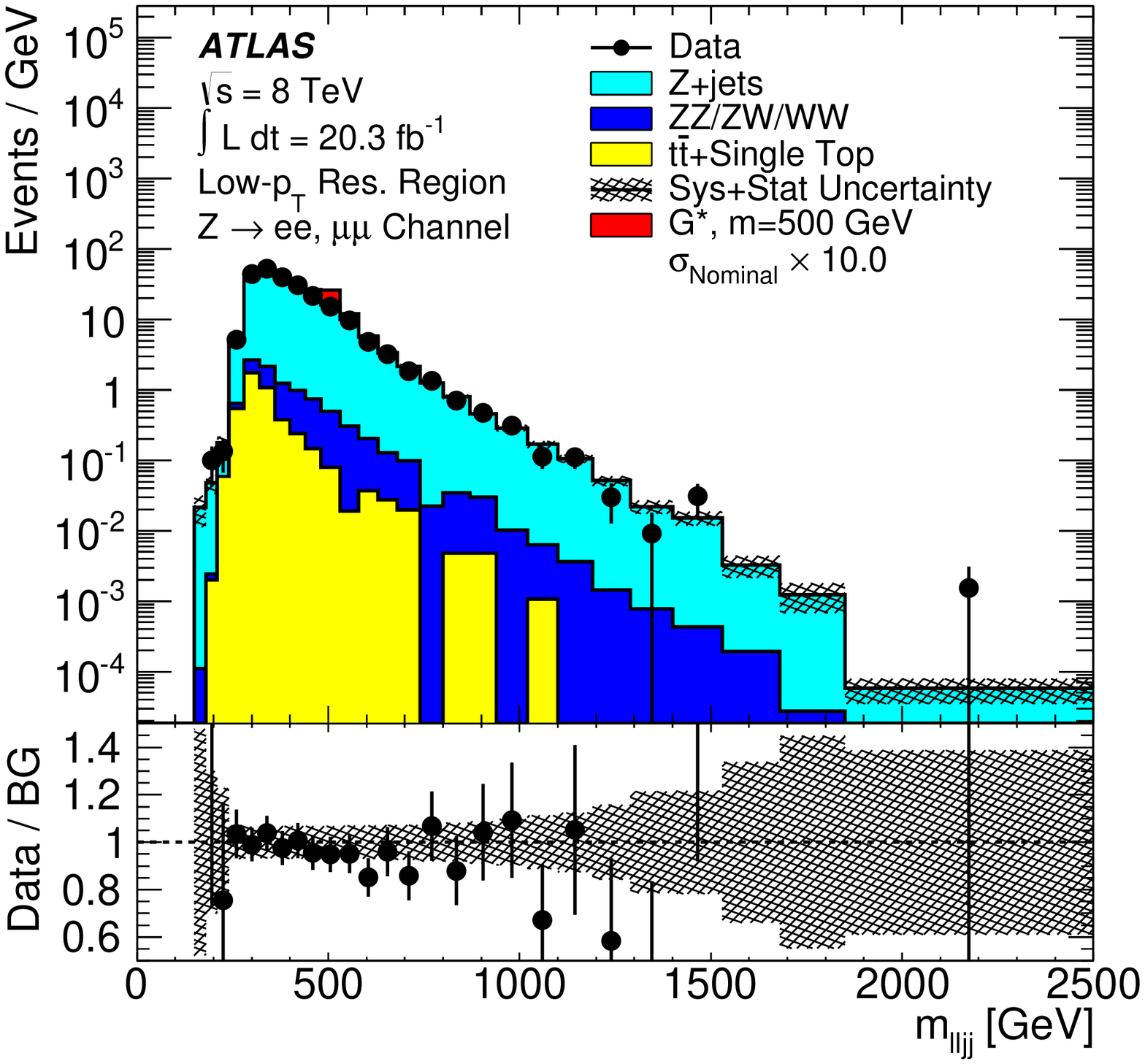} \\
\includegraphics[width=0.8\columnwidth]{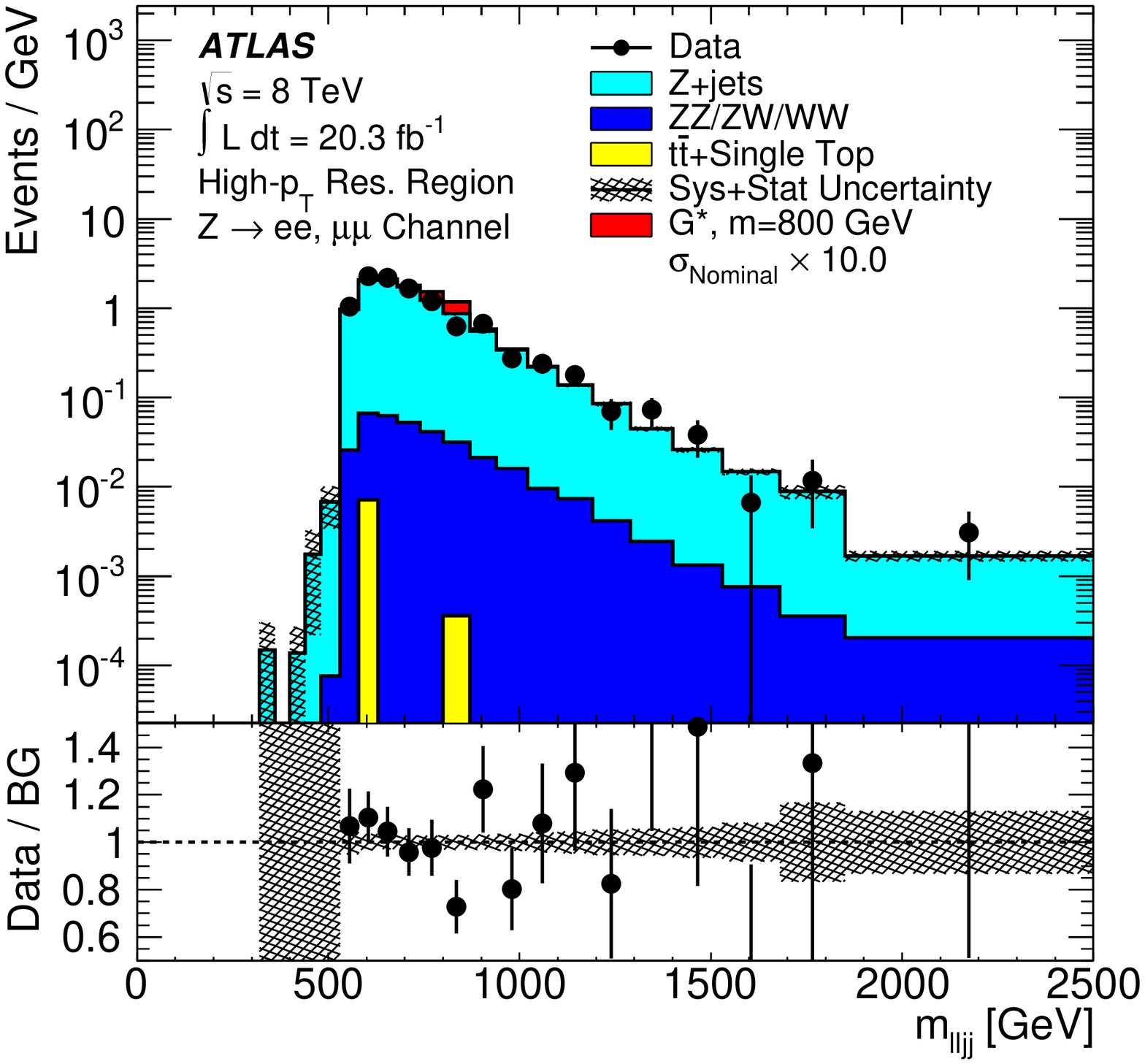} \\
\includegraphics[width=0.8\columnwidth]{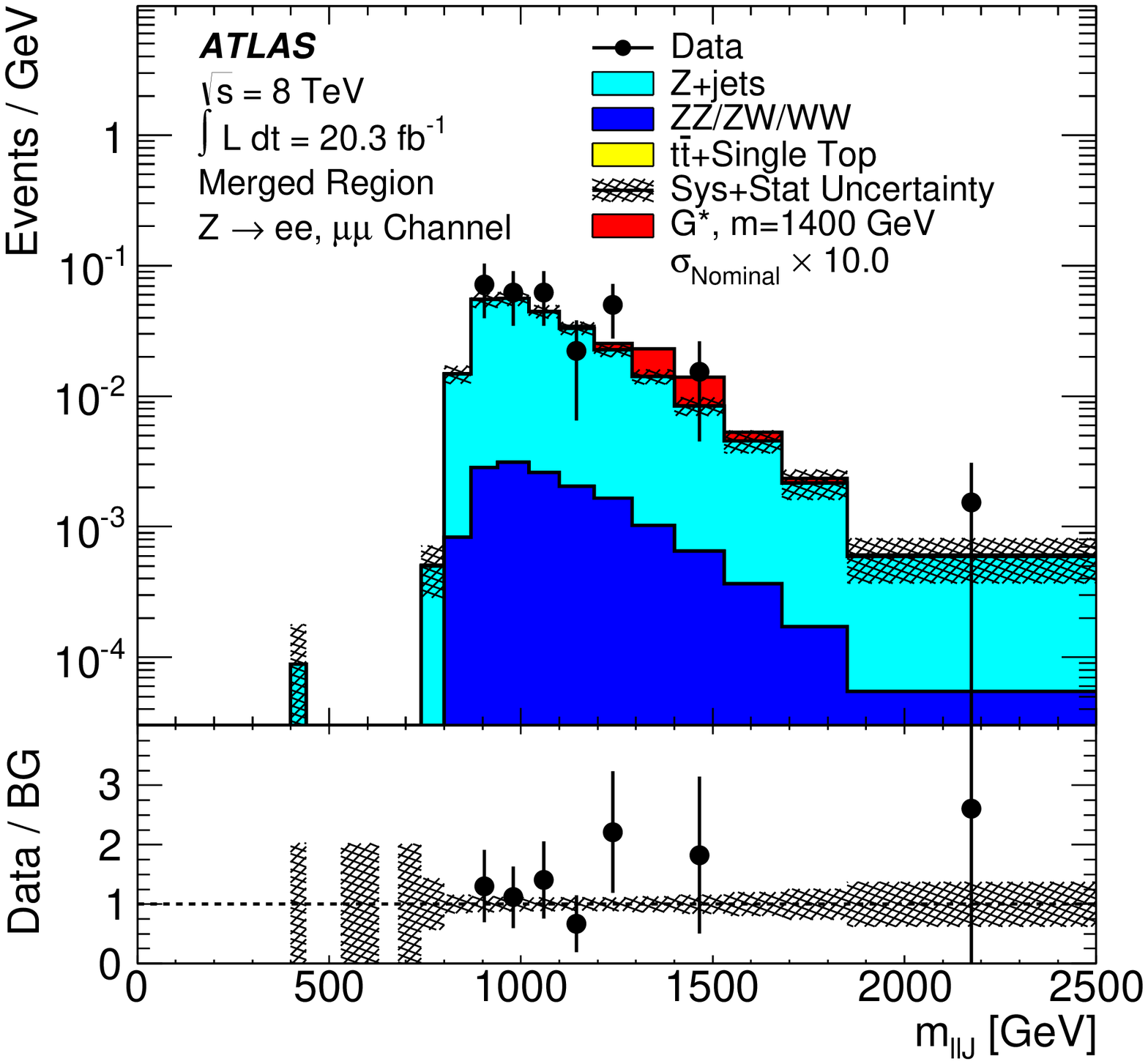}
 \caption{Reconstructed $\ellell jj$ or $\ellell J$ mass distributions in the data and for 
  background after all the selection cuts are applied 
  in the three kinematic regions referred to as the LR (top), HR (middle) and MR (bottom) in the text. 
  The shaded regions show the full background uncertainty obtained by adding statistical and systematic 
  uncertainties in quadrature, including the constraints on the background from the data control 
  regions and before the fit to the data in the signal regions (cf. unconstrained case in Table~\ref{tab:yield}).  
  Also shown are the $\Graviton$ signal yields expected for masses of 500, 800 and 1400~GeV 
  with the production cross sections scaled by a factor 10.}
  \label{fig:mass}
\end{figure}

\subsection{Systematic uncertainties}
\label{sec:systematics}
 The main systematic uncertainty on the $m_{\ell\ell jj/J}$ spectrum comes from the uncertainty in the
$\Zjet$ background modelling. The normalization uncertainty of the $\Zjet$ background is estimated from 
the relative difference between the normalization corrections derived from the nominal control region ($m_{jj/J}<70$~GeV or $>110$~GeV)
and either the lower or higher mass region, taking the larger of the two as an estimate of the systematic uncertainty. 
If the resulting uncertainty is smaller than the statistical uncertainty of the normalization correction from the nominal control region, the latter is used as the systematic uncertainty.
The uncertainty of the shape correction is estimated from the uncertainty on the slope parameter 
of the linear fit and is treated as uncorrelated with respect to the normalization uncertainty.
The combined normalization and shape uncertainties vary as a function of $m_{\ell\ell jj/J}$ 
and range from 
6\% to 9\% in the LR, 
2\% to 8\% in the HR, and 
11\% to 47\% in the MR.  
For all simulated samples, detector performance-related systematic uncertainties 
including the small-$R$ jet energy scale and resolution, 
large-$R$ jet energy, mass and momentum-balance scales and resolutions, the lepton reconstruction and identification efficiencies, 
and lepton momentum scales and resolutions are also considered.
The large-$R$ jet energy and mass scale uncertainties are evaluated by comparing the ratio of 
calorimeter-based to track-based measurements in dijet data and simulated events, and are validated using a data sample of 
high-$\pT$ $W$ bosons produced in association with jets.
A Kolmogorov--Smirnov (KS) test~\cite{b-KStest} is then performed between the nominal and systematically varied distributions
for a given systematic uncertainty source to determine if it has a sizeable effect on the shape of background and signal estimations.
Only significant systematic effects are retained in the analysis by requiring a KS probability of less then 10\%.
For the normalization, if the event yield changes by more than half the statistical uncertainty of the nominal yield, 
the systematic uncertainty is included.
 
Uncertainties on signal acceptance due to PDF sets, renormalization and factorization 
scale choices, initial- and final-state gluon radiation (ISR/FSR) modelling, and LHC beam energy 
uncertainty are also considered. 
The PDF uncertainties are estimated by taking the acceptance difference between CTEQL1 and MSTW2008LO PDFs and adding it
in quadrature to the difference between MSTW2008LO error sets.
The uncertainties due to the scale and ISR/FSR modelling are 
estimated by varying relevant parameters in \pythia8 by a factor of 2.0 and 0.5.
The beam energy systematic uncertainty is assessed with simulation by varying the beam energy
within the measured uncertainty of 0.66\%~\cite{b-beamenergy}, leading to at most a 1\% effect on acceptance. 
The dominant uncertainty comes from ISR/FSR modelling and is approximately 5\%.

\section{Results}
\label{sec:result}
The invariant mass of the diboson system is reconstructed from the $\ellell jj$ 
or $\ellell J$ system.
The reconstructed $m_{\ell\ell jj/J}$ distributions for data and simulated background 
events in the three signal regions are shown in Fig.~\ref{fig:mass} for the combined electron 
and muon channels. Good agreement is observed between the
data and the background predictions, with $p$-values\footnote{The $p$-value is the probability that the background can produce a fluctuation greater than or equal to the excess observed in data.} ranging from 0.98 to 0.10, 
and the results are presented as 95\% confidence level upper limits on the 
production cross section times branching ratio for the $\Graviton$ and $\Wprime$ models.
The upper limits are determined using the $CL_S$ modified frequentist formalism~\cite{b-read2002}
with a profile likelihood test statistic~\cite{b-cowan2011}. The test statistic is evaluated with 
a maximum likelihood fit of signal models and background predictions to the reconstructed 
$m_{\ell\ell jj/J}$ spectra shown in Fig.~\ref{fig:mass}.
Systematic uncertainties and their correlations are taken into account as nuisance parameters
with Gaussian constraints.
The likelihood fit, which takes into account correlations between the systematic 
uncertainties, is performed for signal pole masses ranging between 300--850~GeV 
for the LR, 550--1800~GeV for the HR and 800--2000~GeV for the MR. Overlapping regions
are fit simultaneously.

Fig.~\ref{fig:limit} shows 95\% CL upper limits on the production cross section
times branching fraction into $\Z\Z$ or $\Z\W$ as a function of the resonance pole mass. 
The theoretical predictions for the EGM $W'$ 
and the bulk RS $\Graviton$ with two different values of the coupling constant, 
shown in the figure, allow the extraction of observed (expected) lower 
mass limits of \WprimeMassLimitObs{} (\WprimeMassLimitExp)~GeV for the $W'$,
and \GravitonMassLimitObs{} and \GravitonHalfKMplMassLimitObs{} (\GravitonMassLimitExp{} and 
\GravitonHalfKMplMassLimitExp{})~GeV for the $\Graviton$ 
with $k/\bar{M}_{\rm Pl}=1.0$ and 0.5 respectively.
The most powerful search regions are the LR for masses below 550~GeV, the HR from 500 to 850~GeV and the MR for higher masses.

\begin{figure}[H]\centering
\includegraphics[width=0.9\columnwidth]{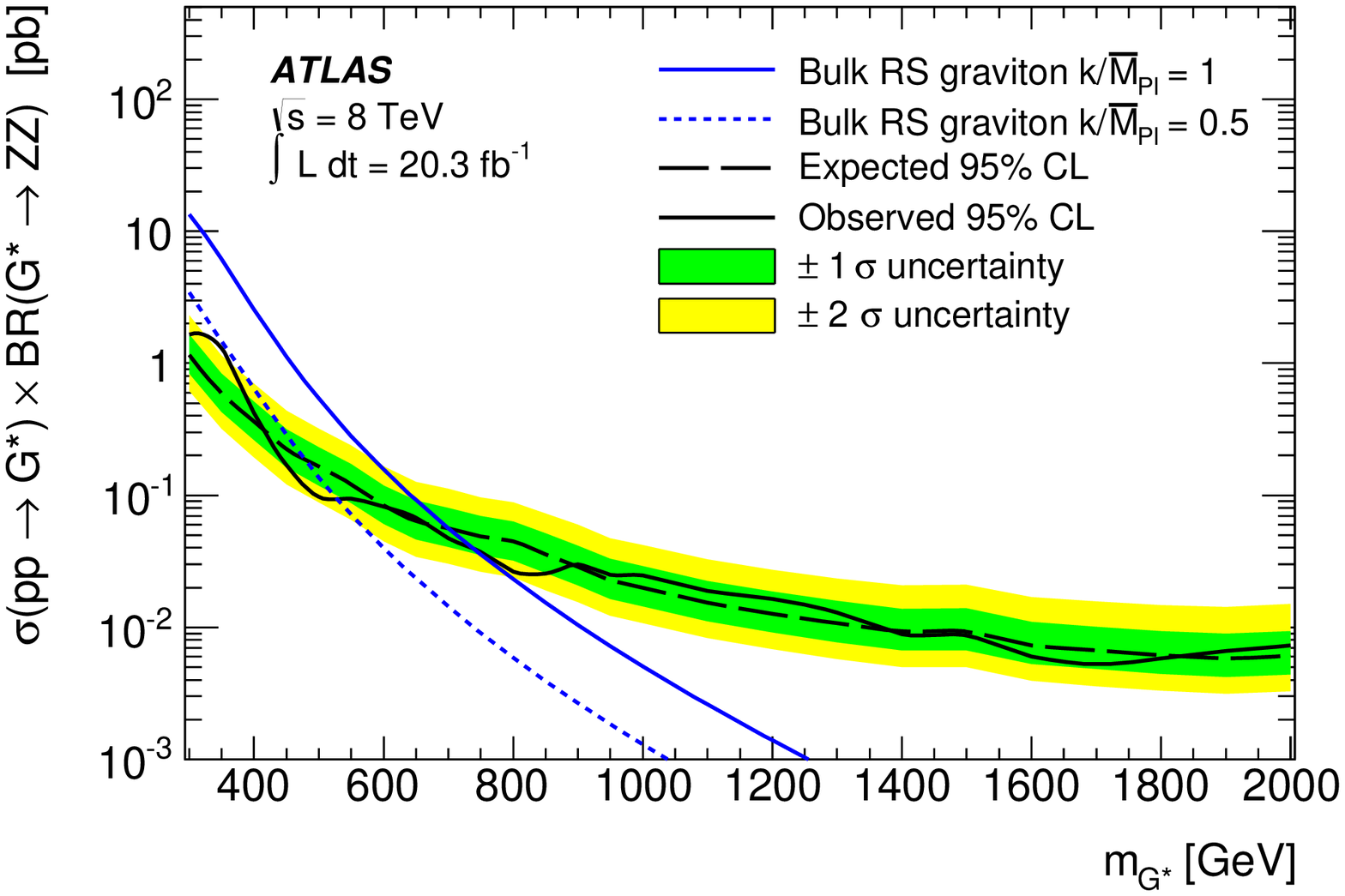}
\includegraphics[width=0.9\columnwidth]{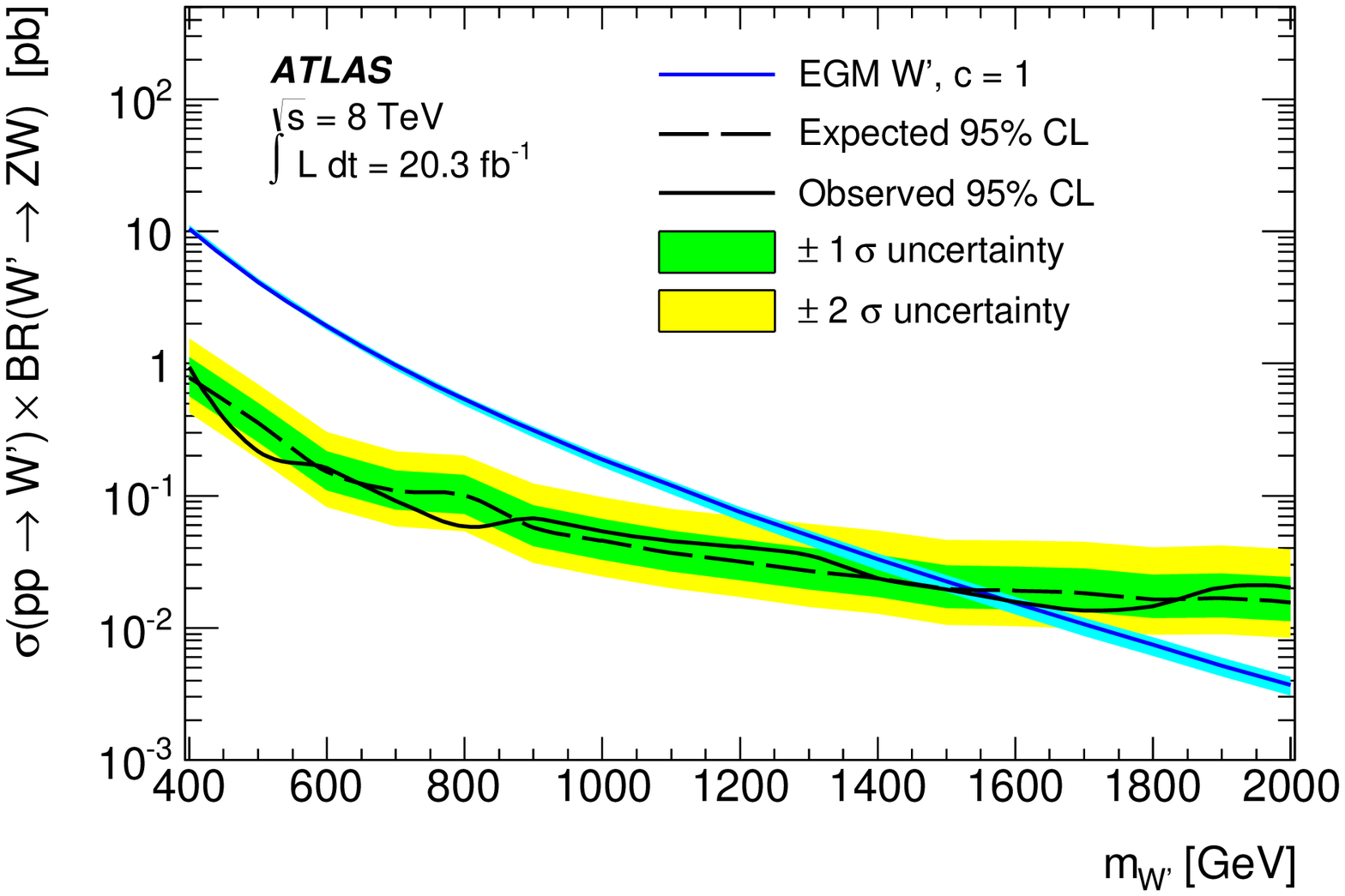}
 \caption{Observed and expected 95\% CL upper limits on the cross section times branching fraction as a 
function of the resonance pole mass for the $\Graviton$ (top) and EGM $\Wprime$ (bottom). 
The LO (NNLO) theoretical cross sections for $\Graviton$ (EGM $\Wprime$) 
production with $k/\bar{M}_{\rm Pl}=0.5$ and 1.0 ($c=1$) are also shown. The band around the $\Wprime$
cross section represents the theoretical uncertainty on the NNLO calculation. The inner and outer bands on the
expected limits represent $\pm1\sigma$ and $\pm2\sigma$ variations respectively.}
\label{fig:limit}
\end{figure}

\section{Conclusion}
\label{sec:conclusion}
In summary, a search for narrow, heavy resonances produced in $pp$ collisions 
and decaying to diboson final states  
at the Large Hadron Collider has been performed. The data sample analysed,
corresponding to an integrated luminosity of $\lumiNoDecimal$~fb$^{-1}$ at $\sqrt{s}=8$~TeV, was recorded  
with the ATLAS detector. No significant excess over the Standard Model background expectation 
was found.  Upper limits on the
production cross section times branching ratio and mass exclusion limits are derived for $\Wprime$ bosons in the theoretical 
framework of an Extended Gauge Model and for gravitons in warped extra dimensions in the 
context of the bulk Randall--Sundrum model. The results represent a significant improvement over previously 
reported limits by ATLAS~\cite{ATLAS_ZZ4l_2l2j}
due to increased $pp$ collision energy and data set size as well as the development of new 
techniques to analyse heavily boosted decays of bosons. 

\begin{acknowledgements}
\input{Acknowledgements}

\end{acknowledgements}

\newpage
\onecolumn
\clearpage 
\input{atlas_authlist} 

\end{document}

%% file: definitions.tex
\def\lumi{20.3} 
\def\lumiNoDecimal{20} 
\def\lumiRelErr{2.8} 

\def\GravitonHalfKMplMassLimitObs{540} 
\def\GravitonMassLimitObs{740} 
\def\WprimeMassLimitObs{1590} 
\def\GravitonHalfKMplMassLimitExp{490} 
\def\GravitonMassLimitExp{700} 
\def\WprimeMassLimitExp{1540} 

\def\GravitonXsec400GeV{3.73e-01} 
\def\GravitonXsec800GeV{3.37e-03} 
\def\GravitonXsec1600GeV{2.24e-05} 
\def\WprimeXsec400GeV{4.74e-01} 
\def\WprimeXsec800GeV{2.433e-02} 
\def\WprimeXsec1600GeV{6.98e-04} 

\def\dzero{D0}

\chardef\letterchar=11
\chardef\otherchar=12
\chardef\eolinechar=5
\def\Z{\ensuremath{Z }}
\def\W{\ensuremath{W }}
\def\calchep{{\sc CalcHEP }}
\def\sherpa{{\sc Sherpa }}
\def\HERWIG{{\sc Herwig }}
\def\JIMMY{{\sc Jimmy }}
\def\PYTHIA{{\sc Pythia }}
\def\ZWPROD{{\sc ZWprod }}

\usepackage{xspace}
\def\WW{\ensuremath{\W\W}\xspace}
\def\WZ{\ensuremath{\W\Zzero}\xspace}

\def\ZZ{\ensuremath{\Zzero\Zzero}\xspace}

\def\pythia{{\sc Pythia}\xspace}

\def\mcatnlo{MC@NLO\xspace}

\def\geant4{{\sc Geant4}\xspace}

\def\ptsp1{\ensuremath{p_T} }

\def\ns{\ifmmode {\mathrm{\ ns}}\else
                   \textrm{ns}\fi}%
\newcount\hrs\newcount\minu\newcount\temptime
\def\hm{\hrs=\time \divide\hrs by 60 \minu=\time\temptime=\hrs
\multiply\temptime by 60%
\advance\minu by -\temptime
\ifnum\minu<10 \let\zerofill=0\else \let\zerofill=\relax\fi
 \the\hrs:\zerofill\the\minu}
\def\lapprox{\ensuremath{\sim\kern-1em\raise 0.65ex\hbox{$<$}}}
\def\rapprox{\ensuremath{\sim\kern-1em\raise 0.65ex\hbox{$>$}}}

\def\rts {\ensuremath{\sqrt{s}}}

\def\antibar#1{\ensuremath{#1\bar{#1}}}

\def\qqbar{\antibar{q}}

\def\ppbar{\antibar{p}}
\def\Zzero{\ensuremath{Z}}

\def\Zjet{\ensuremath{Z+\mathrm{jets}}}
\def\chic{\ensuremath{\raise.4ex\hbox{$\chi$}_{{c}}}} 
\def\chib{\ensuremath{\raise.4ex\hbox{$\chi$}_{{b}}}} 

\let\psii=\psi  
\def\psi{\ensuremath{\psii}}%

\def\ellell{\ensuremath{\mathrm{\ell\ell}}}%
\def\ggino{\ensuremath{\mathchoice%
      {\displaystyle\raise.4ex\hbox{$\displaystyle\tilde\chi$}}%
         {\textstyle\raise.4ex\hbox{$\textstyle\tilde\chi$}}%
       {\scriptstyle\raise.3ex\hbox{$\scriptstyle\tilde\chi$}}%
 {\scriptscriptstyle\raise.3ex\hbox{$\scriptscriptstyle\tilde\chi$}}}}

\def\chinop{\ensuremath{\mathchoice%
      {\displaystyle\raise.4ex\hbox{$\displaystyle\tilde\chi^+$}}%
         {\textstyle\raise.4ex\hbox{$\textstyle\tilde\chi^+$}}%
       {\scriptstyle\raise.3ex\hbox{$\scriptstyle\tilde\chi^+$}}%
 {\scriptscriptstyle\raise.3ex\hbox{$\scriptscriptstyle\tilde\chi^+$}}}}
\def\chinom{\ensuremath{\mathchoice%
      {\displaystyle\raise.4ex\hbox{$\displaystyle\tilde\chi^-$}}%
         {\textstyle\raise.4ex\hbox{$\textstyle\tilde\chi^-$}}%
       {\scriptstyle\raise.3ex\hbox{$\scriptstyle\tilde\chi^-$}}%
 {\scriptscriptstyle\raise.3ex\hbox{$\scriptscriptstyle\tilde\chi^-$}}}}
\def\chinopm{\ensuremath{\mathchoice%
      {\displaystyle\raise.4ex\hbox{$\displaystyle\tilde\chi^\pm$}}%
         {\textstyle\raise.4ex\hbox{$\textstyle\tilde\chi^\pm$}}%
       {\scriptstyle\raise.3ex\hbox{$\scriptstyle\tilde\chi^\pm$}}%
 {\scriptscriptstyle\raise.3ex\hbox{$\scriptscriptstyle\tilde\chi^\pm$}}}}
\def\chinomp{\ensuremath{\mathchoice%
      {\displaystyle\raise.4ex\hbox{$\displaystyle\tilde\chi^\mp$}}%
         {\textstyle\raise.4ex\hbox{$\textstyle\tilde\chi^\mp$}}%
       {\scriptstyle\raise.3ex\hbox{$\scriptstyle\tilde\chi^\mp$}}%
 {\scriptscriptstyle\raise.3ex\hbox{$\scriptscriptstyle\tilde\chi^\mp$}}}}

\def\chinoonep{\ensuremath{\mathchoice%
      {\displaystyle\raise.4ex\hbox{$\displaystyle\tilde\chi^+_1$}}%
         {\textstyle\raise.4ex\hbox{$\textstyle\tilde\chi^+_1$}}%
       {\scriptstyle\raise.3ex\hbox{$\scriptstyle\tilde\chi^+_1$}}%
 {\scriptscriptstyle\raise.3ex\hbox{$\scriptscriptstyle\tilde\chi^+_1$}}}}
\def\chinoonem{\ensuremath{\mathchoice%
      {\displaystyle\raise.4ex\hbox{$\displaystyle\tilde\chi^-_1$}}%
         {\textstyle\raise.4ex\hbox{$\textstyle\tilde\chi^-_1$}}%
       {\scriptstyle\raise.3ex\hbox{$\scriptstyle\tilde\chi^-_1$}}%
 {\scriptscriptstyle\raise.3ex\hbox{$\scriptscriptstyle\tilde\chi^-_1$}}}}
\def\chinoonepm{\ensuremath{\mathchoice%
      {\displaystyle\raise.4ex\hbox{$\displaystyle\tilde\chi^\pm_1$}}%
         {\textstyle\raise.4ex\hbox{$\textstyle\tilde\chi^\pm_1$}}%
       {\scriptstyle\raise.3ex\hbox{$\scriptstyle\tilde\chi^\pm_1$}}%
 {\scriptscriptstyle\raise.3ex\hbox{$\scriptscriptstyle\tilde\chi^\pm_1$}}}}

\def\chinotwop{\ensuremath{\mathchoice%
      {\displaystyle\raise.4ex\hbox{$\displaystyle\tilde\chi^+_2$}}%
         {\textstyle\raise.4ex\hbox{$\textstyle\tilde\chi^+_2$}}%
       {\scriptstyle\raise.3ex\hbox{$\scriptstyle\tilde\chi^+_2$}}%
 {\scriptscriptstyle\raise.3ex\hbox{$\scriptscriptstyle\tilde\chi^+_2$}}}}
\def\chinotwom{\ensuremath{\mathchoice%
      {\displaystyle\raise.4ex\hbox{$\displaystyle\tilde\chi^-_2$}}%
         {\textstyle\raise.4ex\hbox{$\textstyle\tilde\chi^-_2$}}%
       {\scriptstyle\raise.3ex\hbox{$\scriptstyle\tilde\chi^-_2$}}%
 {\scriptscriptstyle\raise.3ex\hbox{$\scriptscriptstyle\tilde\chi^-_2$}}}}
\def\chinotwopm{\ensuremath{\mathchoice%
      {\displaystyle\raise.4ex\hbox{$\displaystyle\tilde\chi^\pm_2$}}%
         {\textstyle\raise.4ex\hbox{$\textstyle\tilde\chi^\pm_2$}}%
       {\scriptstyle\raise.3ex\hbox{$\scriptstyle\tilde\chi^\pm_2$}}%
 {\scriptscriptstyle\raise.3ex\hbox{$\scriptscriptstyle\tilde\chi^\pm_2$}}}}

\def\nino{\ensuremath{\mathchoice%
      {\displaystyle\raise.4ex\hbox{$\displaystyle\tilde\chi^0$}}%
         {\textstyle\raise.4ex\hbox{$\textstyle\tilde\chi^0$}}%
       {\scriptstyle\raise.3ex\hbox{$\scriptstyle\tilde\chi^0$}}%
 {\scriptscriptstyle\raise.3ex\hbox{$\scriptscriptstyle\tilde\chi^0$}}}}

\def\ninoone{\ensuremath{\mathchoice%
      {\displaystyle\raise.4ex\hbox{$\displaystyle\tilde\chi^0_1$}}%
         {\textstyle\raise.4ex\hbox{$\textstyle\tilde\chi^0_1$}}%
       {\scriptstyle\raise.3ex\hbox{$\scriptstyle\tilde\chi^0_1$}}%
 {\scriptscriptstyle\raise.3ex\hbox{$\scriptscriptstyle\tilde\chi^0_1$}}}}
\def\ninotwo{\ensuremath{\mathchoice%
      {\displaystyle\raise.4ex\hbox{$\displaystyle\tilde\chi^0_2$}}%
         {\textstyle\raise.4ex\hbox{$\textstyle\tilde\chi^0_2$}}%
       {\scriptstyle\raise.3ex\hbox{$\scriptstyle\tilde\chi^0_2$}}%
 {\scriptscriptstyle\raise.3ex\hbox{$\scriptscriptstyle\tilde\chi^0_2$}}}}
\def\ninothree{\ensuremath{\mathchoice%
      {\displaystyle\raise.4ex\hbox{$\displaystyle\tilde\chi^0_3$}}%
         {\textstyle\raise.4ex\hbox{$\textstyle\tilde\chi^0_3$}}%
       {\scriptstyle\raise.3ex\hbox{$\scriptstyle\tilde\chi^0_3$}}%
 {\scriptscriptstyle\raise.3ex\hbox{$\scriptscriptstyle\tilde\chi^0_3$}}}}
\def\ninofour{\ensuremath{\mathchoice%
      {\displaystyle\raise.4ex\hbox{$\displaystyle\tilde\chi^0_4$}}%
         {\textstyle\raise.4ex\hbox{$\textstyle\tilde\chi^0_4$}}%
       {\scriptstyle\raise.3ex\hbox{$\scriptstyle\tilde\chi^0_4$}}%
 {\scriptscriptstyle\raise.3ex\hbox{$\scriptscriptstyle\tilde\chi^0_4$}}}}

\def\Graviton{\ensuremath{G^{*}}}

\def\Wprime{\ensuremath{W^\prime}}
\def\mWprime{\ensuremath{m_{W^\prime}}}

\let\pii=\pi
\def\pi{\ensuremath{\pii}}%
\let\etaa=\eta
\def\eta{\ensuremath{\etaa}}%
\def\pt{\ensuremath{p_{\mathrm{T}}}} 
\def\pT{\ensuremath{p_{\mathrm{T}}}} 

\def\mW{\ensuremath{m_W}}

\def\TeV{\ifmmode {\mathrm{\ Te\kern -0.1em V}}\else
                   \textrm{Te\kern -0.1em V}\fi}%
\def\GeV{\ifmmode {\mathrm{\ Ge\kern -0.1em V}}\else
                   \textrm{Ge\kern -0.1em V}\fi}%
\def\MeV{\ifmmode {\mathrm{\ Me\kern -0.1em V}}\else
                   \textrm{Me\kern -0.1em V}\fi}%
\def\keV{\ifmmode {\mathrm{\ ke\kern -0.1em V}}\else
                   \textrm{ke\kern -0.1em V}\fi}%
\def\eV{\ifmmode  {\mathrm{\ e\kern -0.1em V}}\else
                   \textrm{e\kern -0.1em V}\fi}%

\def\TeVc{\ifmmode {\mathrm{\ Te\kern -0.1em V}/c}\else
                   {\textrm{Te\kern -0.1em V}/$c$}\fi}%
\def\GeVc{\ifmmode {\mathrm{\ Ge\kern -0.1em V}/c}\else
                   {\textrm{Ge\kern -0.1em V}/$c$}\fi}%
\def\MeVc{\ifmmode {\mathrm{\ Me\kern -0.1em V}/c}\else
                   {\textrm{Me\kern -0.1em V}/$c$}\fi}%
\def\keVc{\ifmmode {\mathrm{\ ke\kern -0.1em V}/c}\else
                   {\textrm{ke\kern -0.1em V}/$c$}\fi}%
\def\eVc{\ifmmode  {\mathrm{\ e\kern -0.1em V}/c}\else
                   {\textrm{e\kern -0.1em V}/$c$}\fi}%

\def\TeVcc{\ifmmode {\mathrm{\ Te\kern -0.1em V}/c^2}\else
                   {\textrm{Te\kern -0.1em V}/$c^2$}\fi}%
\def\GeVcc{\ifmmode {\mathrm{\ Ge\kern -0.1em V}/c^2}\else
                   {\textrm{Ge\kern -0.1em V}/$c^2$}\fi}%
\def\MeVcc{\ifmmode {\mathrm{\ Me\kern -0.1em V}/c^2}\else
                   {\textrm{Me\kern -0.1em V}/$c^2$}\fi}%
\def\keVcc{\ifmmode {\mathrm{\ ke\kern -0.1em V}/c^2}\else
                   {\textrm{ke\kern -0.1em V}/$c^2$}\fi}%
\def\eVcc{\ifmmode  {\mathrm{\ e\kern -0.1em V}/c^2}\else
                   {\textrm{e\kern -0.1em V}/$c^2$}\fi}%

\def\cm{\ifmmode  {\mathrm{\ cm}}\else
                   \textrm{~cm}\fi}%
\def\ifb{\mbox{fb$^{-1}$}}
\newbox\boxsqbox
\newdimen\boxsize\boxsize=1.2ex%
\def\boxop{%
\setbox\boxsqbox=\vbox{\hrule depth0.8pt width0.8\boxsize height0pt%
                       \kern0.8\boxsize
                       \hrule height0.8pt width0.8\boxsize depth0pt}%
           \hbox{%
           \vrule height1.0\boxsize width0.8pt depth0pt%
           \copy\boxsqbox
           \vrule height1.0\boxsize width0.8pt depth0pt\kern1.5pt}}%
\def\pmb#1{\setbox0=\hbox{$#1$}
  \kern-.025em\copy0\kern-1.0\wd0%
  \kern.05em\copy0\kern-1.0\wd0%
  \kern-.025em\raise.0433em\box0}%
\newdimen\dkwidth
\def\dk{%
   \dkwidth=\baselineskip
   {\def\to{\rightarrow}
   \kern 3pt%
   \hbox{%
      \raise 3pt%
      \hbox{%
         \vrule height 0.8\dkwidth width 0.7pt depth0pt%
      }%
      \kern-0.4pt%
      \hbox to 1.5\dkwidth{%
         \rightarrowfill
      }%
   \kern0.6em%
   }}%
}%
\def\eqalign#1{%
   \,
   \vcenter{%
      \openup\jot\m@th
      \ialign{%
         \strut\hfil$\displaystyle{##}$&&$%
         \displaystyle{{}##}$\hfil\crcr#1\crcr%
      }%
   }%
   \,
}%

%% file: Acknowledgements.tex

We thank CERN for the very successful operation of the LHC, as well as the
support staff from our institutions without whom ATLAS could not be
operated efficiently.

We acknowledge the support of ANPCyT, Argentina; YerPhI, Armenia; ARC,
Australia; BMWFW and FWF, Austria; ANAS, Azerbaijan; SSTC, Belarus; CNPq and FAPESP,
Brazil; NSERC, NRC and CFI, Canada; CERN; CONICYT, Chile; CAS, MOST and NSFC,
China; COLCIENCIAS, Colombia; MSMT CR, MPO CR and VSC CR, Czech Republic;
DNRF, DNSRC and Lundbeck Foundation, Denmark; EPLANET, ERC and NSRF, European Union;
IN2P3-CNRS, CEA-DSM/IRFU, France; GNSF, Georgia; BMBF, DFG, HGF, MPG and AvH
Foundation, Germany; GSRT and NSRF, Greece; RGC, Hong Kong SAR, China; ISF, MINERVA, GIF, I-CORE and Benoziyo Center, Israel; INFN, Italy; MEXT and JSPS, Japan; CNRST, Morocco; FOM and NWO, Netherlands; BRF and RCN, Norway; MNiSW and NCN, Poland; GRICES and FCT, Portugal; MNE/IFA, Romania; MES of Russia and ROSATOM, Russian Federation; JINR; MSTD,
Serbia; MSSR, Slovakia; ARRS and MIZ\v{S}, Slovenia; DST/NRF, South Africa;
MINECO, Spain; SRC and Wallenberg Foundation, Sweden; SER, SNSF and Cantons of
Bern and Geneva, Switzerland; NSC, Taiwan; TAEK, Turkey; STFC, the Royal
Society and Leverhulme Trust, United Kingdom; DOE and NSF, United States of
America.

The crucial computing support from all WLCG partners is acknowledged
gratefully, in particular from CERN and the ATLAS Tier-1 facilities at
TRIUMF (Canada), NDGF (Denmark, Norway, Sweden), CC-IN2P3 (France),
KIT/GridKA (Germany), INFN-CNAF (Italy), NL-T1 (Netherlands), PIC (Spain),
ASGC (Taiwan), RAL (UK) and BNL (USA) and in the Tier-2 facilities
worldwide.

%% file: atlas_authlist.tex
\begin{flushleft}
{\Large The ATLAS Collaboration}

\bigskip

G.~Aad$^{\rm 84}$,
B.~Abbott$^{\rm 112}$,
J.~Abdallah$^{\rm 152}$,
S.~Abdel~Khalek$^{\rm 116}$,
O.~Abdinov$^{\rm 11}$,
R.~Aben$^{\rm 106}$,
B.~Abi$^{\rm 113}$,
M.~Abolins$^{\rm 89}$,
O.S.~AbouZeid$^{\rm 159}$,
H.~Abramowicz$^{\rm 154}$
% ...
\bigskip
\\
% ...
$^{11}$ Institute of Physics, Azerbaijan Academy of Sciences, Baku, Azerbaijan\\
% ...
$^{84}$ CPPM, Aix-Marseille Universit{\'e} and CNRS/IN2P3, Marseille, France\\
% ...
$^{89}$ Department of Physics and Astronomy, Michigan State University, East Lansing MI, United States of America\\
% ...
$^{106}$ Nikhef National Institute for Subatomic Physics and University of Amsterdam, Amsterdam, Netherlands\\
% ...
$^{112}$ Homer L. Dodge Department of Physics and Astronomy, University of Oklahoma, Norman OK, United States of America\\
$^{113}$ Department of Physics, Oklahoma State University, Stillwater OK, United States of America\\
% ...
$^{116}$ LAL, Universit{\'e} Paris-Sud and CNRS/IN2P3, Orsay, France\\
% ...
$^{152}$ Institute of Physics, Academia Sinica, Taipei, Taiwan\\
% ...
$^{154}$ Raymond and Beverly Sackler School of Physics and Astronomy, Tel Aviv University, Tel Aviv, Israel\\
% ...
$^{159}$ Department of Physics, University of Toronto, Toronto ON, Canada\\
% ...
\end{flushleft}
